\documentclass[twocolumn,aps,prl,superscriptaddress]{revtex4-1}

\usepackage{epsf}
\usepackage{graphicx}
\usepackage{sidecap}
\usepackage{array}
\usepackage{amsmath}
\usepackage{amssymb}
\usepackage{dcolumn}
\usepackage{epstopdf}
\usepackage{bm}
\usepackage{color}

\def \beq {\begin{equation}}
\def \eeq {\end{equation}}
\newcommand{\sn}{Sn$_{x}$NbSe$_{2-\delta}$}
\newcommand{\pb}{PbTaSe$_{2}$}
\newcommand{\snnb}{SnNbSe$_{2}$}

\newcommand{\su}{Sn$_{0.13}$NbSe$_{1.70}$}
\newcommand{\sd}{Sn$_{0.14}$NbSe$_{1.71}$}
\newcommand{\st}{Sn$_{0.15}$NbSe$_{1.69}$}
\newcommand{\sq}{Sn$_{0.11}$NbSe$_{1.60}$}

\begin{document}
\title{Unusual upper critical fields of the topological nodal-line semimetal candidate {\sn}}
\author{Riffat Munir}
\author{K A M Hasan~Siddiquee}
\author{Charuni Dissanayake}
\affiliation{Department of Physics, University of Central Florida, Orlando, Florida 32816, USA}
\author{Xinzhe Hu}
\author{Yasumasa Takano} 
\affiliation{Department of Physics, University of Florida, Gainesville, Florida 32611, USA}
\author{Eun Sang Choi}
\affiliation{National High Magnetic Field Laboratory, Florida State University, Tallahassee, Florida 32816, USA}
\author{Yasuyuki Nakajima$^{\ast}$}
\affiliation {Department of Physics, University of Central Florida, Orlando, Florida 32816, USA}

\begin{abstract}
  We report superconductivity in {\sn}, a topological nodal-line semimetal candidate with a noncentrosymmetric crystal structure. The superconducting transition temperature $T_{c}$ of this compound is extremely sensitive to Sn concentration $x$ and Se deficiency $\delta$, 5.0 K for {\su} and 8.6 K for {\sd} and {\st}. In all samples, the temperature dependence of the upper critical field $H_{c2}(T)$ differs from the prediction of the Werthamer-Helfand-Hohenberg theory. While the zero-temperature value of the in-plane upper critical field of {\sn} with the higher $T_{c}$ is lower than the Pauli paramagnetic limit $H_{P}$, that of the lower $T_{c}$ sample exceeds $H_{P}$ by a factor of $\sim$2. Our observations suggest that odd-parity contribution dominates the superconducting gap function of {\sn}, and it can be fine-tuned by the Sn concentration and Se deficiency.
  
\end{abstract}

\date{\today}
\maketitle

%\section{Introduction}

Topological superconductors, characterized by a topologically nontrivial gapped state in the bulk with gapless surface states \cite{sato17}, have attracted great interest because of potential applications to topologically protected quantum computing \cite{kitae03,nayak08}. Such nontrivial gapped states can be stabilized by odd-parity Cooper pairing, occasionally realized in noncentrosymmetric superconductors with strong spin-orbit coupling. The lack of inversion symmetry in the crystal structures allows supersosition of spin singlet (even parity) and spin triplet (odd parity) in superconducting gap functions. This mixing of parity can be fine-tuned by the spin-orbit coupling strength, as observed in the crossover from even to odd-parity pairing states in Li$_{2}$(Pd,Pt)$_{3}$Bi as the Pt concentration is varied \cite{harad12}. Hence, the search for noncentrosymmetric superconductors with strong spin-orbit coupling serves as one of the routes toward the realization of topological superconductivity.

The noncentrosymmetric ABSe$_{2}$ (A=Sn or Pb and B=Nb or Ta) family is a promising candidate for a topological superconductor. Recent experimental studies have revealed superconductivity in {\pb} at $T_{c}=3.8$ K \cite{ali14,long16}, with a fully opened superconducting gap probed by various measurements, including specific heat \cite{zhang16}, thermal conductivity \cite{wang16}, and penetration depth measurements \cite{pang16}. Moreover, angle-resolved photoemission spectroscopy has identified this materials as a topological nodal-line semimetal with drumhead surface states \cite{bian16,chang16}. Although the parity of the superconducting pairing state is yet to be determined, the lack of inversion symmetry that along with the nontrivial topological band structure can induce topological superconductivity in this system in {\pb}.

In addition to this compound, PbNbSe$_{2}$, {\snnb}, and PbTaSe$_{2}$ are predicted by {\it ab-initio} calculations to be superconductors with nontrivial topological nodal lines in the band structure \cite{chen16a}. According to the calculations, the superconducting transition temperatures, $T_{c}$, of these materials are expected to be higher than that of {\pb}, and notably, the predicted $T_{c}$ of {\snnb} is 7 K, suggestive of a possible realization of topological superconductivity in this material with the highest $T_{c}$, compared with known bulk topological superconductor candidates, such as half Heusler compounds YPtBi ($T_{c}$ = 0.6 K) \cite{butch11} and YPdBi ($T_{c}$ = 1.6 K) \cite{nakaj15a}, In-doped SnTe ($T_{c}$ = 1.2 K) \cite{sasak12}, metal-intercalated Bi$_{2}$Se$_{3}$ ($T_{c}$ = 3.2 K) \cite{hor10a,krien11,liu15}, and $\beta$-Bi$_{2}$Pd ($T_{c}$ = 5.4 K) \cite{imai12,liu20}. The higher $T_{c}$ can be advantageous to detect Majorana bound sates with the energy spacing $\Delta^2/E_{F}$, where $\Delta$ and $E_{F}$ are the superconducting gap and Fermi energy, respectively, and can eliminate one bottleneck in the exploration for topological superconductivity \cite{beena15}.

We here report unusual superconductivity in {\sn}, which retains the noncentrosymmetric crystal structure of stoichiometric ABSe$_{2}$. The superconducting transition temperature of {\sn} varies with Sn concentration and Se deficiency---up to 8.6 K, the highest among bulk topological superconductor candidates known to date. The measured upper critical fields of {\sn} cannot be described by the Werthamer-Helfand-Hohenberg (WHH) theory for conventional type-II superconductors \cite{werth66}, and the zero-temperature value of the upper critical field for {\su} exceeds the Pauli paramagnetic limit by a factor of $\sim$2. The unusual temperature dependence of the upper critical fields and the enhancement beyond the Pauli paramagnetic limit suggest that an odd-parity component dominates the superconducting gap function of {\sn}, indicating that this compound is a promising candidate for realizing topological superconductivity.

% \section{Experimental Method}

Single crystals of {\sn} were grown by using a self-flux method. A mixture of NbSe$_{2}$ and Sn in the ratio of NbSe$_2$:Sn=1:2$-$6 was sealed in a quartz tube, heated up to 900$^{\circ}$C, kept for 2$-$3 days, and slowly cooled down to 600$^{\circ}$C. The excess of molten Sn flux was removed by centrifuging. The typical size of obtained flake-like crystals is about 1$\times$1$\times$0.3 mm$^{3}$.

\begin{figure}
\centering
\includegraphics[width=8cm]{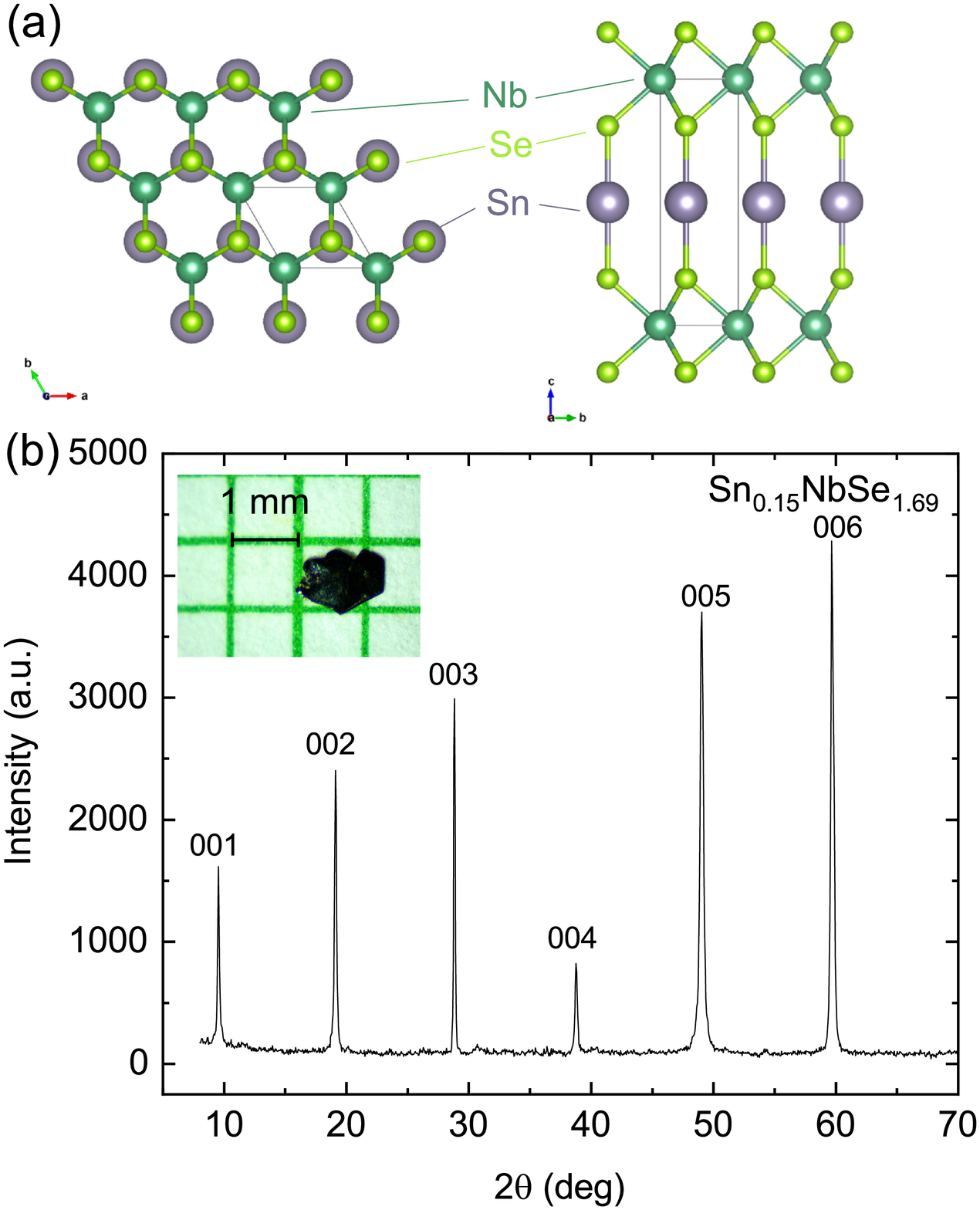}
\caption{{\bf Crystal structure and x-ray diffraction pattern of {\sn}.} (a) Top view (left panel) and side view (right panel) of the crystal structure of {\snnb} with the space group $P\bar 6$m2 \cite{momma11}. (b) X-ray diffraction pattern for single crystal {\st} measured with Cu K$_{\alpha}$ radiation. We only observe (00$\ell$) Bragg peaks from the $ab$ plane. The obtained lattice constant $c$ is 9.2976(12) {\AA}. Inset: Optical image of a {\sn} single crystal. The typical crystal size is $\sim 1\times1\times 0.3$ mm$^{3}$.}
\end{figure}

%\section{Results and Discussion}

The noncentrosymmetric crystal structure of {\sn} with the space group $P\bar 6$m was confirmed by single-crystal x-ray diffraction. As shown in fig.1b, we observed only (00$\ell$) reflections, which yielded the lattice constant $c=$ 9.2976(12) {\AA} close to the reported value of 9.30 {\AA} for polycrystalline {\snnb} \cite{genti79}. The atomic ratio of the crystals was found to be Sn:Nb:Se = 0.11$-$0.15:1:1.60$-$1.74 with x-ray fluorescence spectroscopy, suggestive of slightly intercalated Sn and deficiency of Se. Although the Sn concentrations in our samples are only 11\%$-$15\% of the stoichiometric value, the identified crystal structure is the same as that of ABSe$_{2}$ and different from the 2H-NbSe$_{2}$-type structure of Sn-intercalated NbSe$_{2}$ with $x$ up to 0.04 \cite{naik19}.

We reveal superconductivity in {\sn} with resistivity measurements from 300 K down to 2 K. The resistivity at $T=$ 300 K is about 700~$\mu\Omega$ cm. Upon cooling, the resistivity shows metallic behavior, followed by a superconducting transition at low temperatures (fig.1). The superconducting transition temperatures $T_{c}$, determined by the midpoint of resistive transitions, are 5.0 K for {\su} and 8.6 K for {\sd} and {\st} (fig.1 inset). In contrast, no superconductivity has been observed down to 1.5 K in polycrystalline Sn$_{2/3}$NbSe$_{2}$ \cite{karne75}. The normal-state resistivity of {\sn} right above $T_{c}$ is much higher than that of {\pb}. Because of this high residual resistivity, the residual resistivity ratio RRR = $\rho$(300 K)/$\rho$($T_{c}$) of {\sn} is 2.3$-$3.0, 30$-$50 times as small as that of single-crystal {\pb} \cite{zhang16,long16,sanka17}. Interestingly, despite the high residual resistivity, or the low RRR, the $T_{c}$ of 8.6 K not only surpasses the theoretical prediction of 7 K \cite{chen16a} but is to date the highest superconducting transition temperature among known topological superconductor candidates.

\begin{figure}
\centering
\includegraphics[width=8.5cm]{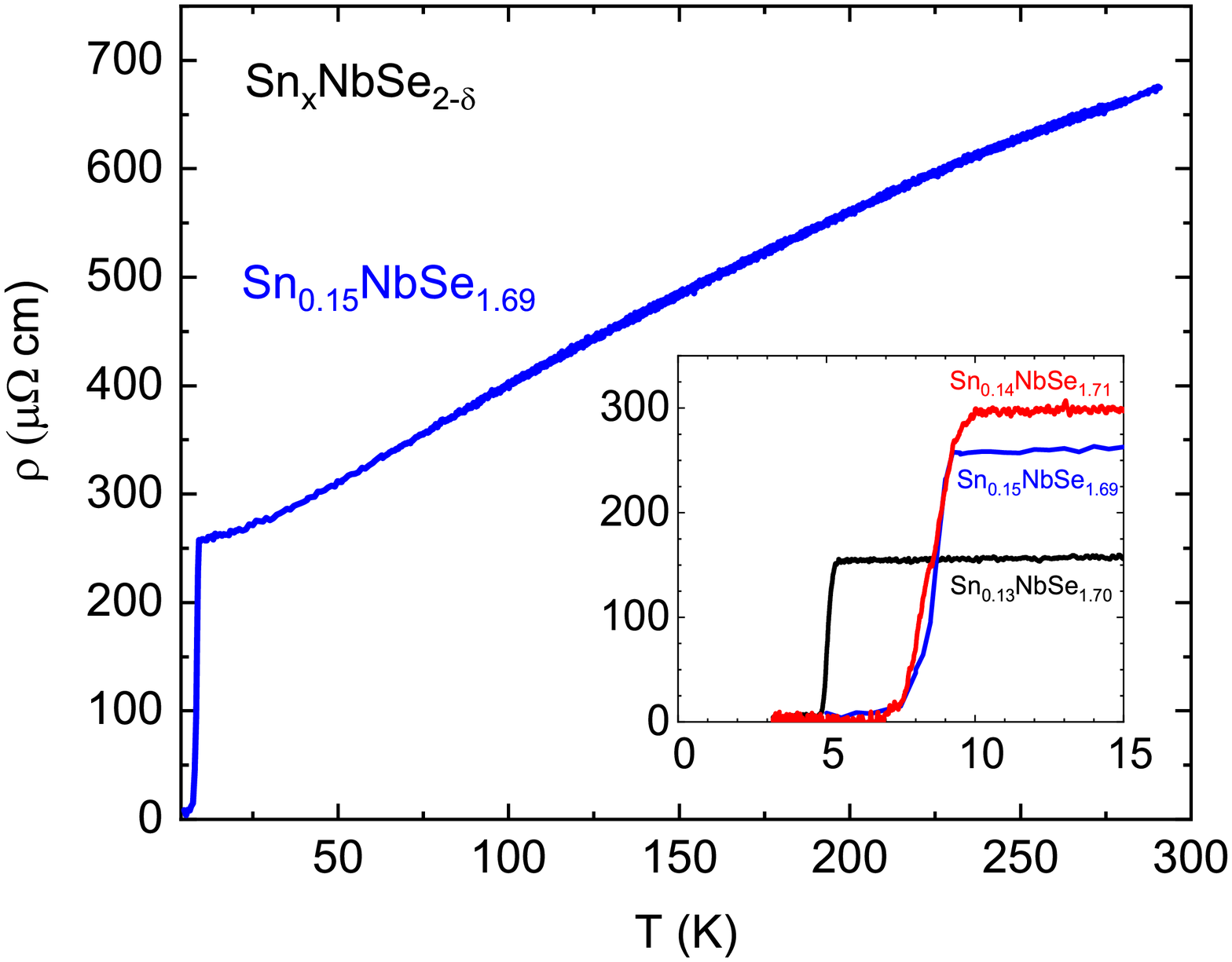}
\caption{{\bf Temperature dependence of the resistivities of {\sn}}. Overall temperature dependence of the resistivity of {\st}. Preceded by metallic behavior, {\sn} undergoes a superconducting transition at low temperatures. The residual resistivity ratio $RRR$ = $\rho$(300 K)/$\rho$($T_{c}$) is 2.3$-$3.0, suggestive of abundant disorder. Inset: Low temperature resistivities of {\sn}, indicating superconducting transitions at 5.0 K for {\su} and 8.6 K for {\sd} and {\st}.}
\end{figure}

% MR

To determine the upper critical fields of {\sn}, we measured the resistivity in a 35 T resistive magnet equipped with a $^{3}$He cryostat at the National High Magnetic Field Laboratory (NHMFL) in Tallahassee, Florida. We show magnetoresistance of {\sn} in two different magnetic field configurations: $H\parallel I \parallel ab$ (in-plane configuration) and $H\perp ab, I\parallel ab$ (out-of-plane configuration). For both field configurations and in both {\su} and {\sd}, the magnetoresistance is indiscernible even at an applied magnetic field of 35 T (figs.3a$-$d), indicative of short mean free paths of charge carriers due to disorder caused most likely by slightly intercalated Sn and deficiency of Se. On the other hand, we find a notable difference in the anisotropy of upper critical fields, $\Gamma = H_{c2}^{ab}/H_{c2}^{c}$, between these two samples. In the out-of-plane configuration for {\su} (fig.3b), the upper critical field parallel to the $c$ axis, $H_{c2}^{c}$, defined by the midpoints of sharp resistive transitions, is 5 T at 0.3 K. In the in-plane configuration for {\su} (fig.3a), the transition widths are broad, and the upper critical field parallel to the $ab$ plane, $H_{c2}^{ab}$, is extremely high, extracted to be 15 T at 0.3 K. These values of $H_{c2}$ yield a large anisotropy of the upper critical field, $\Gamma = 3$, for {\su}. In contrast, we observe similar transition widths with very similar $H_{c2}$ in both field configurations for {\sd}, suggestive of a nearly isotropic upper critical field with $\Gamma = 1.1$ (figs.3c and d). In the isostructural compound PbTaSe$_{2}$, the anisotropy of $H_{c2}$ is measured to be 11.6 \cite{long16}, much larger than the observed anisotropy in {\sn}, although the two compounds share similar electronic band structures \cite{chen16a}.

\begin{figure}
	
\includegraphics[width=8.5cm]{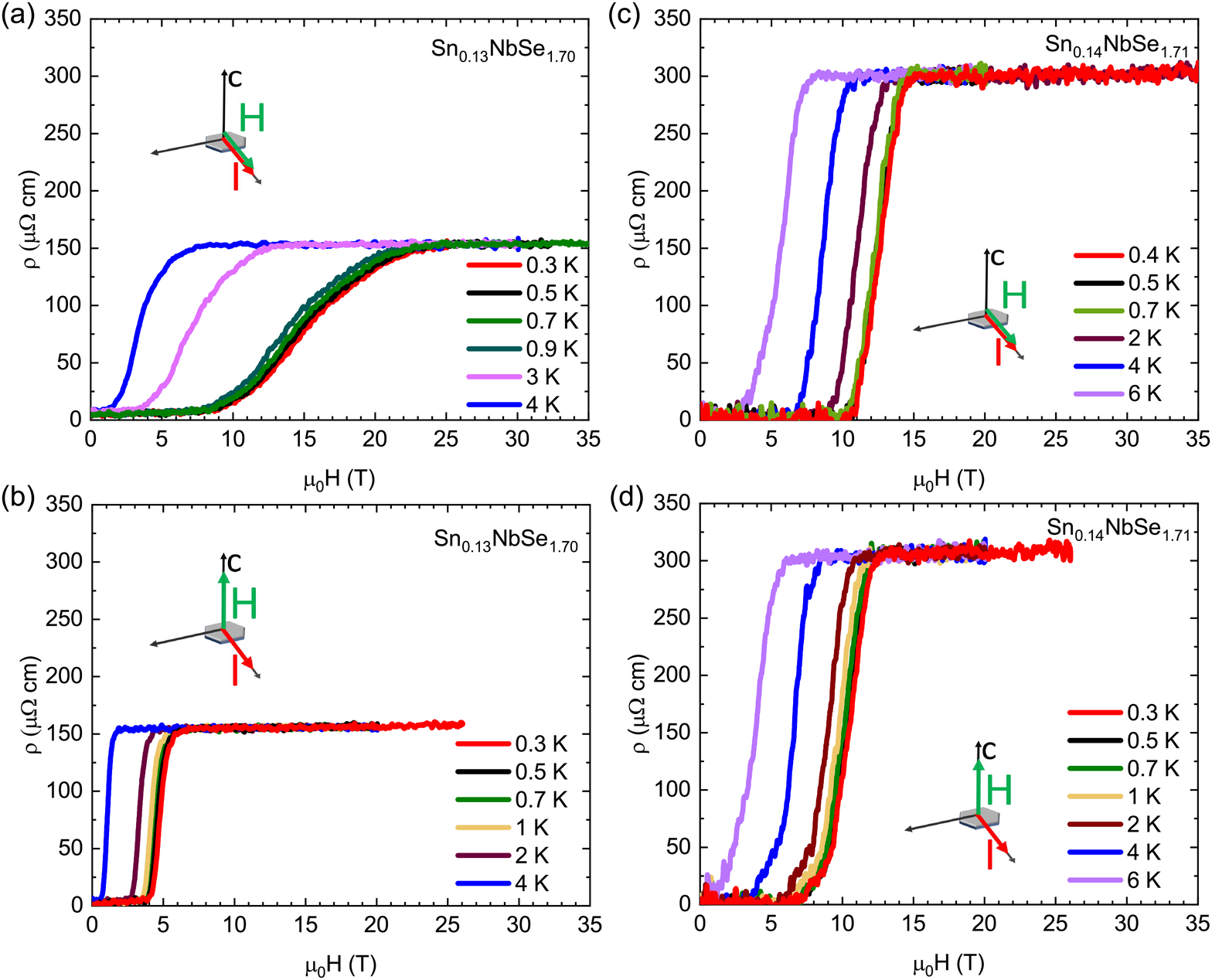}
\caption{{\bf Magnetotransport in {\sn}}. Resistivity of {\su} as a function of magnetic fields parallel to (a) the $ab$ plane and (b) the $c$ axis at several temperatures. Electrical current is applied in the $ab$ plane. While the resistive transitions for $H\parallel c$ are sharp, the transitions for $H\parallel ab$ are broad. The upper critical fields obtained by the midpoints of the resistive transitions show high anisotropy $\Gamma = H_{c2}^{ab}/H_{c2}^{c} =$ 3 in  {\su}. Resistivity of {\sd} as a function of magnetic fields parallel to (c) the $ab$ plane and (d) the $c$ axis at various temperatures. Despite the layered structure of the compound, the upper critical field is nearly isotropic, $\Gamma =$ 1.1, in {\sd}.}
\end{figure}

\begin{figure*}  
\includegraphics[width= 15cm]{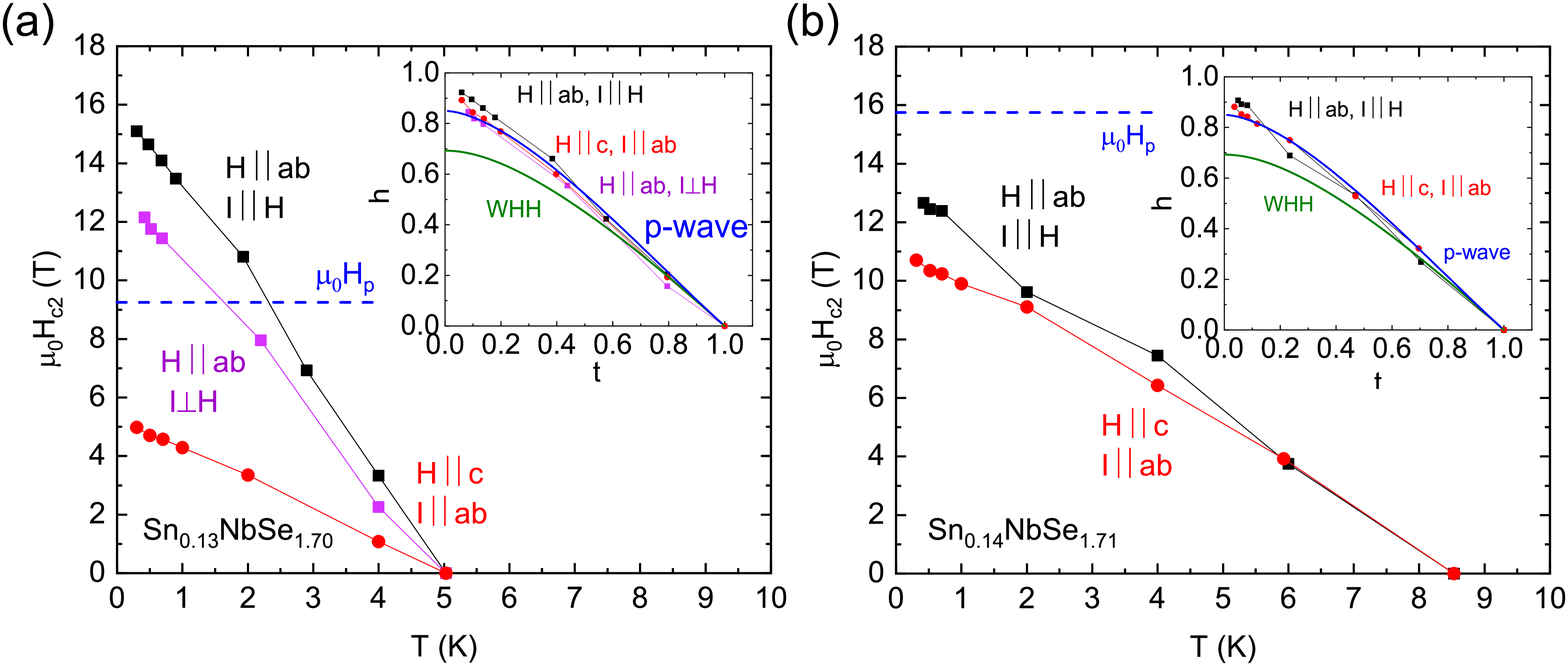}
\caption{{\bf Temperature dependence and anisotropy of the upper critical field of {\sn}.} (a) Temperature dependence of upper critical field determined by the midpoint of resistive transitions in figs.3 and 6 for {\su}. Regardless of the field orientation, the upper critical field increases linearly with decreasing temperature down to 0.3 K. The zero-temperature values of in-plane upper critical fields $\mu_{0} H_{c2}^{ab}(0)$ for $H \parallel I$ and $H \perp I$ exceed the Pauli paramagnetic limit $\mu_{0}H_{P} =1.84 T_{c}=$ 9.3 T, indicated by a blue dashed line. The zero-temperature value of out-of-plane $\mu_{0} H_{c2}^{c}(0)$ is lower than $\mu_{0}H_{P}$. Inset: Normalized upper critical field $h = H_{c2}/(-T_{c}dH_{c2}/dT|_{T=Tc})$ as a function of normalized temperature $t = T/T_{c}$ for {\su}. For all the field orientations, the normalized upper critical field clearly deviates from the WHH model at low temperatures and is close to the calculation for the polar $p$-wave state \cite{schar80}. (b) Temperature dependence of the upper critical field determined by the midpoint of resistive transitions in fig.3 for {\sd}. The upper critical fields for $H\parallel ab$ and $H \parallel c$ exhibit linear-in-temperature behavior down to 0.3 K. The zero-temperature values of in-plane upper critical fields $\mu_{0} H_{c2}^{ab}(0)$ for $H\parallel I$ and $H \perp I$ are lower than the Pauli paramagnetic limit $\mu_{0}H_{P} =$ 15.6 T, indicated by a blue dashed line. Inset: Normalized upper critical field $h$ as a function of $t$ for {\sd}. The normalized upper critical fields for both $H\parallel ab$ and $H \parallel c$ depart from the WHH model at low temperatures and are close to the calculation for the polar $p$-wave state \cite{schar80}.}
\end{figure*}

The upper critical fields of {\sn} exhibit quite unusual behavior, as shown in fig.4. With decreasing temperature starting from $T_{c}$, $H_{c2}$ increases linearly in all the field configurations for both {\su} and {\sd}. The initial slopes of the upper critical fields parallel to the $ab$ plane for $H\parallel I$, $dH_{c2}/dT|_{T=T_{c}}$, are $-$3.2 T/K for {\su} and $-$1.5 T/K for {\sd}, extremely larger than $-$0.19 K/T for {\pb} for $H\parallel ab$ \cite{zhang16}. The observed temperature dependence of $H_{c2}$ for {\sn} completely differs from those of conventional type-II superconductors. In conventional type-II superconductors, upper critical fields due to the orbital effect are explicated by the Werthamer-Helfand-Hohenberg (WHH) theory \cite{werth66}. In the WHH theory, the zero-temperature value of upper critical field can be written as $H_{c2}(0)=-\alpha T_{c} dH_{c2}/dT |_{T=T_{c}}$, where $\alpha$ is 0.69 for the dirty limit and 0.72 for the clean limit \cite{werth66}. However, as shown in the insets to figs.4a and b, the normalized upper critical field $h= H_{c2}/(-T_{c}dH_{c2}/dT|_{T=T_{c}})$ strikingly deviates from the WHH values and is close to the calculations for the polar $p$-wave state \cite{schar80}. The extracted $\alpha$ for {\sn} is $\sim 1.0$.

Equally unusual in {\sn} are the magnitudes of the upper critical fields. The zero temperature value of the in-plane upper critical field $\mu_{0}H_{c2}^{ab}(0)$ for {\su} is $\sim$16 T, exceeding the Pauli paramagnetic limit $\mu_{0}H_{P} = 1.84 T_{c} =$ 9.3 T obtained from a simple estimation within the weak-coupling BCS theory with $2\Delta$ = 3.5$k_{B}T_{c}$ (fig.4a) \cite{clogs62}. The persistence of superconductivity beyond the Pauli paramagnetic limit suggests dominant odd-parity contribution in the superconducting gap function,as found in noncentrosymmetric heavy fermions superconductor CePt$_{2}$Si \cite{bauer04} and metal-intercalated Bi$_{2}$Se$_{2}$ \cite{bay12}.

In {\sd}, $\mu_{0}H_{c2}^{ab}(0)$ is 13 T, comparable to but slightly lower than $\mu_{0}H_{P}=$ 15.6 T enhanced by the higher $T_{c}$ of 8.6 K (fig.4b). The zero temperature values of out-of-plane upper critical fields $\mu_{0}H_{c2}^{c}(0)$ of {\sn} are lower than the Pauli paramagnetic limit. Utilizing the Ginzburg–Landau relation, $\mu_{0}H_{c2}^{c}(0)=\phi_{0}/2\pi\xi_{ab}^{2}$ and $\mu_{0}H_{c2}^{c}(0)=\phi_{0}/2\pi \xi_{c}\xi_{ab}$, where $\phi_{0}$ is the magnetic flux quantum, we can estimate the coherence lengths: $\xi_{ab}=$ 15 nm and $\xi_{c}=$ 4.5 nm for {\su} and $\xi_{ab}=$ 5.0 nm and $\xi_{c}=$ 6.0 nm for {\sd}.

Although our observations suggest the presence of considerable odd-parity contribution to the pairing states at least in {\su} with $T_{c}$ of 5.0 K, its superconductivity is robust against scattering due to nonmagnetic disorder. In general, nonmagnetic scattering strongly suppresses unconventional superconductivity \cite{macke98}, including odd-parity pairing states, as described by the Abrikosov-Gorkov model \cite{abrik61}. The Abrikosov-Gorkov model shows that $T_{c}$ of unconventional superconductors are dramatically suppressed to zero by nonmagnetic scattering, as the mean free path $\ell$ becomes comparable to the coherence length, $\xi \sim \ell$. To estimate the mean free paths in {\sn}, we exploit the Hall effect, along with electrical resistivity. Hall resistivity $\rho_{yx}$ of {\sq} is positive and perfectly linear in magnetic fields in the normal state above $T_{c}$ (fig.5), indicative of a dominant contribution to the charge transport from a large cylindrical hole band, around the $\Gamma$ point, predicted by theoretical calculations \cite{chen16a,xu19}. The slope of the Hall resistivity gives the Hall coefficient $R_{H}=5.3\times$ 10$^{-3}$ cm$^{3}$/C, hence the carrier density $n= 1.2\times 10^{21}$ cm$^{-3}$. Assuming a single cylindrical hole band, we can extract mean free paths by using $\ell = (\hbar/\rho e^{2})\sqrt{2\pi c/n}$, where $c$ is the lattice constant along the $c$ axis. The extracted mean free paths at $T_{c}$ are $\ell$ = 5.9 nm for {\su} and 3.1 nm for {\sd}. Surprisingly, these values are much smaller than the coherence lengths in the $ab$ plane, intuitively inconsistent with the possible odd-parity pairing state.

\begin{figure}  
\includegraphics[width= 8cm]{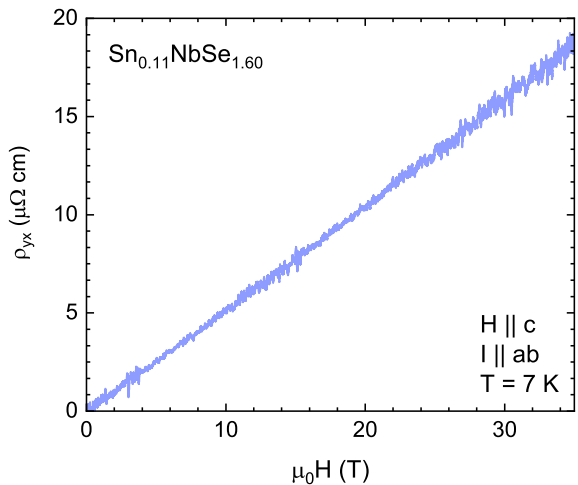}
\caption{{\bf Field dependence of the Hall resistivity of {\sq} at 7 K ($H\parallel c, I\parallel ab$).} The Hall resistivity $\rho_{yx}$ is linear in $H$ with the positive sign indicating that the hole band dominates the transport in {\sn}. The extracted Hall coefficient $R_{H}$ from the slope of $\rho_{yx}$ is 5.3 $\times$ 10$^{-3}$ cm$^{3}$/C, yielding a carrier density of $1.2\times 10^{21}$ cm$^{-3}$.}
\end{figure}

However, there is a precedence in which odd-parity superconductivity was found to be invulnerble to nonmagnetic disorder---metal-intercalated Bi$_{2}$Se$_{3}$ \cite{sasak11,krien11,bay12}. In this topological superconductor, the robustness of the odd-parity superconductivity is thought to be due to strong spin-orbit coupling \cite{micha12,nagai14}. Similarly, nonnegligible spin-orbit coupling due to Sn, a relatively heavy element, may be protecting superconductivity in {\sn} against strong nonmagnetic scattering. The shared features of superconductivity in the two compounds---the upper critical fields beyond the Pauli paramagnetic limit and the robust superconductivity against disorder---strongly suggest that the odd-parity superconductivity is realized in {\sn}.

\begin{figure}  
\includegraphics[width= 8cm]{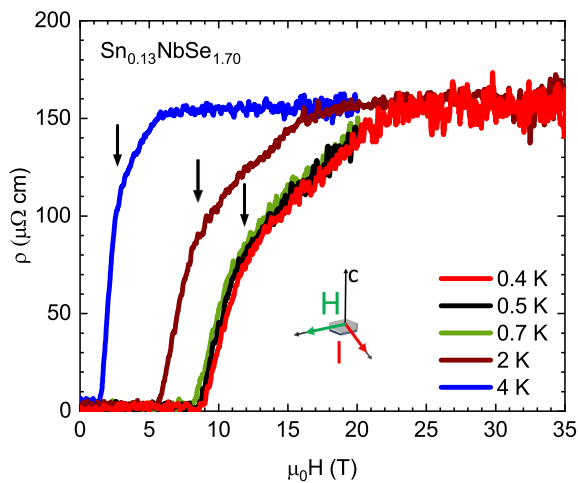}
\caption{{\bf Transverse magnetoresistance of {\su} in magnetic fields parallel to the $ab$ plane.} Resistivity of {\su} as a function of in-plane magnetic field perpendicular to the applied current in the $ab$ plane. The resistive transitions are broad, and show kinks indicated by arrows, associated with vortex motion due to the Lorentz force. The extracted $H_{c2}^{ab}$ for $H\perp I$ is slightly lower than $H_{c2}^{ab}$ for $H\parallel I$, as shown in fig.4a.}
\end{figure}

%nematicity
We briefly comment on a possible nematic superconducting state in the {\pb} family. Nematic superconductivity, accompanied by a rotational symmetry breaking in the superconducting state, is observed in metal-intercalated Bi$_{2}$Se$_{3}$, evidencing topological superconductivity in the system \cite{matan16,pan16,smyli17,yonez17}. Similarly, in {\pb}, soft point-contact spectroscopy \cite{le20} elucidates the rotational symmetry breaking in superconducting properties, possibly associated with nematic superconductivity. In {\su}, we observe clear anisotropy in in-plane upper critical fields for two different configurations, $H\parallel I$ and $H\perp I$, as shown in fig.4a. However, the measured anisotropy can be attributed to the presence/absence of flux flow due to the Lorentz force as observed in MgB$_{2}$ \cite{shi03}, masking intrinsic twofold symmetry associated with electronic nematicity in this system, if present. Indeed, we observe kinks in the resistive transition associated with vortex motion due to the Lorentz force only in the $H\perp I$ configuration as shown in fig.6 \cite{zhu10a}. To determine the intrinsic in-plane anisotropy, further experimental investigations will be required.

In summary, we have grown single crystals of {\sn} with a noncentrosymmetric crystal structure, and found superconductivity with the highest $T_{c}$, 8.6 K among known topological superconductor candidates. The upper critical field of {\su}, with $T_{c}=$ 5 K, exceeds the Pauli paramagnetic limit, suggestive of a sizable contribution of odd-parity paring in the superconductivity. The possible odd-parity pairing component, a prerequisite for topological superconductivity, makes this system a promising material for further study.

\begin{acknowledgments}
This work was supported by the start-up fund from the University of Central Florida. H.S and Y.N. are supported by NSF CAREER DMR-1944975, and X.H. and Y.T. by the NHMFL UCGP program. The NHMFL is supported by the National Science Foundation through NSF/DMR-1644779 and the State of Florida.
\end{acknowledgments}


\begin{thebibliography}{44}
\expandafter\ifx\csname natexlab\endcsname\relax\def\natexlab#1{#1}\fi
\expandafter\ifx\csname bibnamefont\endcsname\relax
  \def\bibnamefont#1{#1}\fi
\expandafter\ifx\csname bibfnamefont\endcsname\relax
  \def\bibfnamefont#1{#1}\fi
\expandafter\ifx\csname citenamefont\endcsname\relax
  \def\citenamefont#1{#1}\fi
\expandafter\ifx\csname url\endcsname\relax
  \def\url#1{\texttt{#1}}\fi
\expandafter\ifx\csname urlprefix\endcsname\relax\def\urlprefix{URL }\fi
\providecommand{\bibinfo}[2]{#2}
\providecommand{\eprint}[2][]{\url{#2}}

\bibitem[{\citenamefont{Sato and Ando}(2017)}]{sato17}
\bibinfo{author}{\bibfnamefont{M.}~\bibnamefont{Sato}} \bibnamefont{and}
  \bibinfo{author}{\bibfnamefont{Y.}~\bibnamefont{Ando}},
  \bibinfo{journal}{Reports on Progress in Physics}
  \textbf{\bibinfo{volume}{80}}, \bibinfo{pages}{076501}
  (\bibinfo{year}{2017}).

\bibitem[{\citenamefont{Kitaev}(2003)}]{kitae03}
\bibinfo{author}{\bibfnamefont{A.}~\bibnamefont{Kitaev}},
  \bibinfo{journal}{Annals of Physics} \textbf{\bibinfo{volume}{303}},
  \bibinfo{pages}{2 } (\bibinfo{year}{2003}).

\bibitem[{\citenamefont{Nayak et~al.}(2008)\citenamefont{Nayak, Simon, Stern,
  Freedman, and Das~Sarma}}]{nayak08}
\bibinfo{author}{\bibfnamefont{C.}~\bibnamefont{Nayak}},
  \bibinfo{author}{\bibfnamefont{S.~H.} \bibnamefont{Simon}},
  \bibinfo{author}{\bibfnamefont{A.}~\bibnamefont{Stern}},
  \bibinfo{author}{\bibfnamefont{M.}~\bibnamefont{Freedman}}, \bibnamefont{and}
  \bibinfo{author}{\bibfnamefont{S.}~\bibnamefont{Das~Sarma}},
  \bibinfo{journal}{Rev. Mod. Phys.} \textbf{\bibinfo{volume}{80}},
  \bibinfo{pages}{1083} (\bibinfo{year}{2008}).

\bibitem[{\citenamefont{Harada et~al.}(2012)\citenamefont{Harada, Zhou, Yao,
  Inada, and Zheng}}]{harad12}
\bibinfo{author}{\bibfnamefont{S.}~\bibnamefont{Harada}},
  \bibinfo{author}{\bibfnamefont{J.~J.} \bibnamefont{Zhou}},
  \bibinfo{author}{\bibfnamefont{Y.~G.} \bibnamefont{Yao}},
  \bibinfo{author}{\bibfnamefont{Y.}~\bibnamefont{Inada}}, \bibnamefont{and}
  \bibinfo{author}{\bibfnamefont{G.-q.} \bibnamefont{Zheng}},
  \bibinfo{journal}{Phys. Rev. B} \textbf{\bibinfo{volume}{86}},
  \bibinfo{pages}{220502} (\bibinfo{year}{2012}).

\bibitem[{\citenamefont{Ali et~al.}(2014)\citenamefont{Ali, Gibson, Klimczuk,
  and Cava}}]{ali14}
\bibinfo{author}{\bibfnamefont{M.~N.} \bibnamefont{Ali}},
  \bibinfo{author}{\bibfnamefont{Q.~D.} \bibnamefont{Gibson}},
  \bibinfo{author}{\bibfnamefont{T.}~\bibnamefont{Klimczuk}}, \bibnamefont{and}
  \bibinfo{author}{\bibfnamefont{R.~J.} \bibnamefont{Cava}},
  \bibinfo{journal}{Phys. Rev. B} \textbf{\bibinfo{volume}{89}},
  \bibinfo{pages}{020505} (\bibinfo{year}{2014}).

\bibitem[{\citenamefont{Long et~al.}(2016)\citenamefont{Long, Zhao, Wang, Yang,
  Li, Zi, Ren, Ren, and Chen}}]{long16}
\bibinfo{author}{\bibfnamefont{Y.-J.} \bibnamefont{Long}},
  \bibinfo{author}{\bibfnamefont{L.-X.} \bibnamefont{Zhao}},
  \bibinfo{author}{\bibfnamefont{P.-P.} \bibnamefont{Wang}},
  \bibinfo{author}{\bibfnamefont{H.-X.} \bibnamefont{Yang}},
  \bibinfo{author}{\bibfnamefont{J.-Q.} \bibnamefont{Li}},
  \bibinfo{author}{\bibfnamefont{H.}~\bibnamefont{Zi}},
  \bibinfo{author}{\bibfnamefont{Z.-A.} \bibnamefont{Ren}},
  \bibinfo{author}{\bibfnamefont{C.}~\bibnamefont{Ren}}, \bibnamefont{and}
  \bibinfo{author}{\bibfnamefont{G.-F.} \bibnamefont{Chen}},
  \bibinfo{journal}{Chinese Physics Letters} \textbf{\bibinfo{volume}{33}},
  \bibinfo{pages}{037401} (\bibinfo{year}{2016}).

\bibitem[{\citenamefont{Zhang et~al.}(2016)\citenamefont{Zhang, Yuan, Bian, Xu,
  Zhang, Hasan, and Jia}}]{zhang16}
\bibinfo{author}{\bibfnamefont{C.-L.} \bibnamefont{Zhang}},
  \bibinfo{author}{\bibfnamefont{Z.}~\bibnamefont{Yuan}},
  \bibinfo{author}{\bibfnamefont{G.}~\bibnamefont{Bian}},
  \bibinfo{author}{\bibfnamefont{S.-Y.} \bibnamefont{Xu}},
  \bibinfo{author}{\bibfnamefont{X.}~\bibnamefont{Zhang}},
  \bibinfo{author}{\bibfnamefont{M.~Z.} \bibnamefont{Hasan}}, \bibnamefont{and}
  \bibinfo{author}{\bibfnamefont{S.}~\bibnamefont{Jia}},
  \bibinfo{journal}{Phys. Rev. B} \textbf{\bibinfo{volume}{93}},
  \bibinfo{pages}{054520} (\bibinfo{year}{2016}).

\bibitem[{\citenamefont{Wang et~al.}(2016)\citenamefont{Wang, Xu, He, Zhang,
  Hong, Cai, Wang, Dong, and Li}}]{wang16}
\bibinfo{author}{\bibfnamefont{M.~X.} \bibnamefont{Wang}},
  \bibinfo{author}{\bibfnamefont{Y.}~\bibnamefont{Xu}},
  \bibinfo{author}{\bibfnamefont{L.~P.} \bibnamefont{He}},
  \bibinfo{author}{\bibfnamefont{J.}~\bibnamefont{Zhang}},
  \bibinfo{author}{\bibfnamefont{X.~C.} \bibnamefont{Hong}},
  \bibinfo{author}{\bibfnamefont{P.~L.} \bibnamefont{Cai}},
  \bibinfo{author}{\bibfnamefont{Z.~B.} \bibnamefont{Wang}},
  \bibinfo{author}{\bibfnamefont{J.~K.} \bibnamefont{Dong}}, \bibnamefont{and}
  \bibinfo{author}{\bibfnamefont{S.~Y.} \bibnamefont{Li}},
  \bibinfo{journal}{Phys. Rev. B} \textbf{\bibinfo{volume}{93}},
  \bibinfo{pages}{020503} (\bibinfo{year}{2016}).

\bibitem[{\citenamefont{Pang et~al.}(2016)\citenamefont{Pang, Smidman, Zhao,
  Wang, Weng, Che, Chen, Lu, Chen, and Yuan}}]{pang16}
\bibinfo{author}{\bibfnamefont{G.~M.} \bibnamefont{Pang}},
  \bibinfo{author}{\bibfnamefont{M.}~\bibnamefont{Smidman}},
  \bibinfo{author}{\bibfnamefont{L.~X.} \bibnamefont{Zhao}},
  \bibinfo{author}{\bibfnamefont{Y.~F.} \bibnamefont{Wang}},
  \bibinfo{author}{\bibfnamefont{Z.~F.} \bibnamefont{Weng}},
  \bibinfo{author}{\bibfnamefont{L.~Q.} \bibnamefont{Che}},
  \bibinfo{author}{\bibfnamefont{Y.}~\bibnamefont{Chen}},
  \bibinfo{author}{\bibfnamefont{X.}~\bibnamefont{Lu}},
  \bibinfo{author}{\bibfnamefont{G.~F.} \bibnamefont{Chen}}, \bibnamefont{and}
  \bibinfo{author}{\bibfnamefont{H.~Q.} \bibnamefont{Yuan}},
  \bibinfo{journal}{Phys. Rev. B} \textbf{\bibinfo{volume}{93}},
  \bibinfo{pages}{060506} (\bibinfo{year}{2016}).

\bibitem[{\citenamefont{Bian et~al.}(2016)\citenamefont{Bian, Chang, Sankar,
  Xu, Zheng, Neupert, Chiu, Huang, Chang, Belopolski et~al.}}]{bian16}
\bibinfo{author}{\bibfnamefont{G.}~\bibnamefont{Bian}},
  \bibinfo{author}{\bibfnamefont{T.-R.} \bibnamefont{Chang}},
  \bibinfo{author}{\bibfnamefont{R.}~\bibnamefont{Sankar}},
  \bibinfo{author}{\bibfnamefont{S.-Y.} \bibnamefont{Xu}},
  \bibinfo{author}{\bibfnamefont{H.}~\bibnamefont{Zheng}},
  \bibinfo{author}{\bibfnamefont{T.}~\bibnamefont{Neupert}},
  \bibinfo{author}{\bibfnamefont{C.-K.} \bibnamefont{Chiu}},
  \bibinfo{author}{\bibfnamefont{S.-M.} \bibnamefont{Huang}},
  \bibinfo{author}{\bibfnamefont{G.}~\bibnamefont{Chang}},
  \bibinfo{author}{\bibfnamefont{I.}~\bibnamefont{Belopolski}},
  \bibnamefont{et~al.}, \bibinfo{journal}{Nature Communications}
  \textbf{\bibinfo{volume}{7}}, \bibinfo{pages}{10556 EP }
  (\bibinfo{year}{2016}).

\bibitem[{\citenamefont{Chang et~al.}(2016)\citenamefont{Chang, Chen, Bian,
  Huang, Zheng, Neupert, Sankar, Xu, Belopolski, Chang et~al.}}]{chang16}
\bibinfo{author}{\bibfnamefont{T.-R.} \bibnamefont{Chang}},
  \bibinfo{author}{\bibfnamefont{P.-J.} \bibnamefont{Chen}},
  \bibinfo{author}{\bibfnamefont{G.}~\bibnamefont{Bian}},
  \bibinfo{author}{\bibfnamefont{S.-M.} \bibnamefont{Huang}},
  \bibinfo{author}{\bibfnamefont{H.}~\bibnamefont{Zheng}},
  \bibinfo{author}{\bibfnamefont{T.}~\bibnamefont{Neupert}},
  \bibinfo{author}{\bibfnamefont{R.}~\bibnamefont{Sankar}},
  \bibinfo{author}{\bibfnamefont{S.-Y.} \bibnamefont{Xu}},
  \bibinfo{author}{\bibfnamefont{I.}~\bibnamefont{Belopolski}},
  \bibinfo{author}{\bibfnamefont{G.}~\bibnamefont{Chang}},
  \bibnamefont{et~al.}, \bibinfo{journal}{Phys. Rev. B}
  \textbf{\bibinfo{volume}{93}}, \bibinfo{pages}{245130}
  (\bibinfo{year}{2016}).

\bibitem[{\citenamefont{Chen et~al.}(2016)\citenamefont{Chen, Chang, and
  Jeng}}]{chen16a}
\bibinfo{author}{\bibfnamefont{P.-J.} \bibnamefont{Chen}},
  \bibinfo{author}{\bibfnamefont{T.-R.} \bibnamefont{Chang}}, \bibnamefont{and}
  \bibinfo{author}{\bibfnamefont{H.-T.} \bibnamefont{Jeng}},
  \bibinfo{journal}{Phys. Rev. B} \textbf{\bibinfo{volume}{94}},
  \bibinfo{pages}{165148} (\bibinfo{year}{2016}).

\bibitem[{\citenamefont{Butch et~al.}(2011)\citenamefont{Butch, Syers,
  Kirshenbaum, Hope, and Paglione}}]{butch11}
\bibinfo{author}{\bibfnamefont{N.~P.} \bibnamefont{Butch}},
  \bibinfo{author}{\bibfnamefont{P.}~\bibnamefont{Syers}},
  \bibinfo{author}{\bibfnamefont{K.}~\bibnamefont{Kirshenbaum}},
  \bibinfo{author}{\bibfnamefont{A.~P.} \bibnamefont{Hope}}, \bibnamefont{and}
  \bibinfo{author}{\bibfnamefont{J.}~\bibnamefont{Paglione}},
  \bibinfo{journal}{Phys. Rev. B} \textbf{\bibinfo{volume}{84}},
  \bibinfo{pages}{220504} (\bibinfo{year}{2011}).

\bibitem[{\citenamefont{Nakajima et~al.}(2015)\citenamefont{Nakajima, Hu,
  Kirshenbaum, Hughes, Syers, Wang, Wang, Wang, Saha, Pratt et~al.}}]{nakaj15a}
\bibinfo{author}{\bibfnamefont{Y.}~\bibnamefont{Nakajima}},
  \bibinfo{author}{\bibfnamefont{R.}~\bibnamefont{Hu}},
  \bibinfo{author}{\bibfnamefont{K.}~\bibnamefont{Kirshenbaum}},
  \bibinfo{author}{\bibfnamefont{A.}~\bibnamefont{Hughes}},
  \bibinfo{author}{\bibfnamefont{P.}~\bibnamefont{Syers}},
  \bibinfo{author}{\bibfnamefont{X.}~\bibnamefont{Wang}},
  \bibinfo{author}{\bibfnamefont{K.}~\bibnamefont{Wang}},
  \bibinfo{author}{\bibfnamefont{R.}~\bibnamefont{Wang}},
  \bibinfo{author}{\bibfnamefont{S.~R.} \bibnamefont{Saha}},
  \bibinfo{author}{\bibfnamefont{D.}~\bibnamefont{Pratt}},
  \bibnamefont{et~al.}, \bibinfo{journal}{Science Advances}
  \textbf{\bibinfo{volume}{1}} (\bibinfo{year}{2015}).

\bibitem[{\citenamefont{Sasaki et~al.}(2012)\citenamefont{Sasaki, Ren, Taskin,
  Segawa, Fu, and Ando}}]{sasak12}
\bibinfo{author}{\bibfnamefont{S.}~\bibnamefont{Sasaki}},
  \bibinfo{author}{\bibfnamefont{Z.}~\bibnamefont{Ren}},
  \bibinfo{author}{\bibfnamefont{A.~A.} \bibnamefont{Taskin}},
  \bibinfo{author}{\bibfnamefont{K.}~\bibnamefont{Segawa}},
  \bibinfo{author}{\bibfnamefont{L.}~\bibnamefont{Fu}}, \bibnamefont{and}
  \bibinfo{author}{\bibfnamefont{Y.}~\bibnamefont{Ando}},
  \bibinfo{journal}{Phys. Rev. Lett.} \textbf{\bibinfo{volume}{109}},
  \bibinfo{pages}{217004} (\bibinfo{year}{2012}).

\bibitem[{\citenamefont{Hor et~al.}(2010)\citenamefont{Hor, Williams,
  Checkelsky, Roushan, Seo, Xu, Zandbergen, Yazdani, Ong, and Cava}}]{hor10a}
\bibinfo{author}{\bibfnamefont{Y.~S.} \bibnamefont{Hor}},
  \bibinfo{author}{\bibfnamefont{A.~J.} \bibnamefont{Williams}},
  \bibinfo{author}{\bibfnamefont{J.~G.} \bibnamefont{Checkelsky}},
  \bibinfo{author}{\bibfnamefont{P.}~\bibnamefont{Roushan}},
  \bibinfo{author}{\bibfnamefont{J.}~\bibnamefont{Seo}},
  \bibinfo{author}{\bibfnamefont{Q.}~\bibnamefont{Xu}},
  \bibinfo{author}{\bibfnamefont{H.~W.} \bibnamefont{Zandbergen}},
  \bibinfo{author}{\bibfnamefont{A.}~\bibnamefont{Yazdani}},
  \bibinfo{author}{\bibfnamefont{N.~P.} \bibnamefont{Ong}}, \bibnamefont{and}
  \bibinfo{author}{\bibfnamefont{R.~J.} \bibnamefont{Cava}},
  \bibinfo{journal}{Phys. Rev. Lett.} \textbf{\bibinfo{volume}{104}},
  \bibinfo{pages}{057001} (\bibinfo{year}{2010}).

\bibitem[{\citenamefont{Kriener et~al.}(2011)\citenamefont{Kriener, Segawa,
  Ren, Sasaki, and Ando}}]{krien11}
\bibinfo{author}{\bibfnamefont{M.}~\bibnamefont{Kriener}},
  \bibinfo{author}{\bibfnamefont{K.}~\bibnamefont{Segawa}},
  \bibinfo{author}{\bibfnamefont{Z.}~\bibnamefont{Ren}},
  \bibinfo{author}{\bibfnamefont{S.}~\bibnamefont{Sasaki}}, \bibnamefont{and}
  \bibinfo{author}{\bibfnamefont{Y.}~\bibnamefont{Ando}},
  \bibinfo{journal}{Phys. Rev. Lett.} \textbf{\bibinfo{volume}{106}},
  \bibinfo{pages}{127004} (\bibinfo{year}{2011}).

\bibitem[{\citenamefont{Liu et~al.}(2015)\citenamefont{Liu, Yao, Shao, Zuo, Pi,
  Tan, Zhang, and Zhang}}]{liu15}
\bibinfo{author}{\bibfnamefont{Z.}~\bibnamefont{Liu}},
  \bibinfo{author}{\bibfnamefont{X.}~\bibnamefont{Yao}},
  \bibinfo{author}{\bibfnamefont{J.}~\bibnamefont{Shao}},
  \bibinfo{author}{\bibfnamefont{M.}~\bibnamefont{Zuo}},
  \bibinfo{author}{\bibfnamefont{L.}~\bibnamefont{Pi}},
  \bibinfo{author}{\bibfnamefont{S.}~\bibnamefont{Tan}},
  \bibinfo{author}{\bibfnamefont{C.}~\bibnamefont{Zhang}}, \bibnamefont{and}
  \bibinfo{author}{\bibfnamefont{Y.}~\bibnamefont{Zhang}},
  \bibinfo{journal}{Journal of the American Chemical Society}
  \textbf{\bibinfo{volume}{137}}, \bibinfo{pages}{10512}
  (\bibinfo{year}{2015}).

\bibitem[{\citenamefont{Imai et~al.}(2012)\citenamefont{Imai, Nabeshima,
  Yoshinaka, Miyatani, Kondo, Komiya, Tsukada, and Maeda}}]{imai12}
\bibinfo{author}{\bibfnamefont{Y.}~\bibnamefont{Imai}},
  \bibinfo{author}{\bibfnamefont{F.}~\bibnamefont{Nabeshima}},
  \bibinfo{author}{\bibfnamefont{T.}~\bibnamefont{Yoshinaka}},
  \bibinfo{author}{\bibfnamefont{K.}~\bibnamefont{Miyatani}},
  \bibinfo{author}{\bibfnamefont{R.}~\bibnamefont{Kondo}},
  \bibinfo{author}{\bibfnamefont{S.}~\bibnamefont{Komiya}},
  \bibinfo{author}{\bibfnamefont{I.}~\bibnamefont{Tsukada}}, \bibnamefont{and}
  \bibinfo{author}{\bibfnamefont{A.}~\bibnamefont{Maeda}},
  \bibinfo{journal}{Journal of the Physical Society of Japan}
  \textbf{\bibinfo{volume}{81}}, \bibinfo{pages}{113708}
  (\bibinfo{year}{2012}).

\bibitem[{\citenamefont{Liu et~al.}(2020)\citenamefont{Liu, Li, Tu, Yin, Sa,
  Zhang, Singh, and Wang}}]{liu20}
\bibinfo{author}{\bibfnamefont{P.-F.} \bibnamefont{Liu}},
  \bibinfo{author}{\bibfnamefont{J.}~\bibnamefont{Li}},
  \bibinfo{author}{\bibfnamefont{X.-H.} \bibnamefont{Tu}},
  \bibinfo{author}{\bibfnamefont{H.}~\bibnamefont{Yin}},
  \bibinfo{author}{\bibfnamefont{B.}~\bibnamefont{Sa}},
  \bibinfo{author}{\bibfnamefont{J.}~\bibnamefont{Zhang}},
  \bibinfo{author}{\bibfnamefont{D.~J.} \bibnamefont{Singh}}, \bibnamefont{and}
  \bibinfo{author}{\bibfnamefont{B.-T.} \bibnamefont{Wang}},
  \bibinfo{journal}{Phys. Rev. B} \textbf{\bibinfo{volume}{102}},
  \bibinfo{pages}{155406} (\bibinfo{year}{2020}).

\bibitem[{\citenamefont{Beenakker}(2015)}]{beena15}
\bibinfo{author}{\bibfnamefont{C.~W.~J.} \bibnamefont{Beenakker}},
  \bibinfo{journal}{Rev. Mod. Phys.} \textbf{\bibinfo{volume}{87}},
  \bibinfo{pages}{1037} (\bibinfo{year}{2015}).

\bibitem[{\citenamefont{Werthamer et~al.}(1966)\citenamefont{Werthamer,
  Helfand, and Hohenberg}}]{werth66}
\bibinfo{author}{\bibfnamefont{N.}~\bibnamefont{Werthamer}},
  \bibinfo{author}{\bibfnamefont{E.}~\bibnamefont{Helfand}}, \bibnamefont{and}
  \bibinfo{author}{\bibfnamefont{P.}~\bibnamefont{Hohenberg}},
  \bibinfo{journal}{Phys. Rev.} \textbf{\bibinfo{volume}{147}},
  \bibinfo{pages}{295} (\bibinfo{year}{1966}).

\bibitem[{\citenamefont{Momma and Izumi}(2011)}]{momma11}
\bibinfo{author}{\bibfnamefont{K.}~\bibnamefont{Momma}} \bibnamefont{and}
  \bibinfo{author}{\bibfnamefont{F.}~\bibnamefont{Izumi}},
  \bibinfo{journal}{Journal of Applied Crystallography}
  \textbf{\bibinfo{volume}{44}}, \bibinfo{pages}{1272} (\bibinfo{year}{2011}).

\bibitem[{\citenamefont{Gentile et~al.}(1979)\citenamefont{Gentile, Driscoll,
  and Hockman}}]{genti79}
\bibinfo{author}{\bibfnamefont{P.}~\bibnamefont{Gentile}},
  \bibinfo{author}{\bibfnamefont{D.}~\bibnamefont{Driscoll}}, \bibnamefont{and}
  \bibinfo{author}{\bibfnamefont{A.}~\bibnamefont{Hockman}},
  \bibinfo{journal}{Inorganica Chimica Acta} \textbf{\bibinfo{volume}{35}},
  \bibinfo{pages}{249 } (\bibinfo{year}{1979}).

\bibitem[{\citenamefont{Naik et~al.}(2019)\citenamefont{Naik, Pradhan, Bhat,
  Behera, Kumar, Samal, and Samal}}]{naik19}
\bibinfo{author}{\bibfnamefont{S.}~\bibnamefont{Naik}},
  \bibinfo{author}{\bibfnamefont{G.~K.} \bibnamefont{Pradhan}},
  \bibinfo{author}{\bibfnamefont{S.~G.} \bibnamefont{Bhat}},
  \bibinfo{author}{\bibfnamefont{B.~C.} \bibnamefont{Behera}},
  \bibinfo{author}{\bibfnamefont{P.~A.} \bibnamefont{Kumar}},
  \bibinfo{author}{\bibfnamefont{S.~L.} \bibnamefont{Samal}}, \bibnamefont{and}
  \bibinfo{author}{\bibfnamefont{D.}~\bibnamefont{Samal}},
  \bibinfo{journal}{Physica C: Superconductivity and its Applications}
  \textbf{\bibinfo{volume}{561}}, \bibinfo{pages}{18 } (\bibinfo{year}{2019}).

\bibitem[{\citenamefont{Karnezos et~al.}(1975)\citenamefont{Karnezos, Welsh,
  and Shafer}}]{karne75}
\bibinfo{author}{\bibfnamefont{N.}~\bibnamefont{Karnezos}},
  \bibinfo{author}{\bibfnamefont{L.~B.} \bibnamefont{Welsh}}, \bibnamefont{and}
  \bibinfo{author}{\bibfnamefont{M.~W.} \bibnamefont{Shafer}},
  \bibinfo{journal}{Phys. Rev. B} \textbf{\bibinfo{volume}{11}},
  \bibinfo{pages}{1808} (\bibinfo{year}{1975}).

\bibitem[{\citenamefont{Sankar et~al.}(2017)\citenamefont{Sankar, Rao,
  Muthuselvam, Chang, Jeng, Murugan, Lee, and Chou}}]{sanka17}
\bibinfo{author}{\bibfnamefont{R.}~\bibnamefont{Sankar}},
  \bibinfo{author}{\bibfnamefont{G.~N.} \bibnamefont{Rao}},
  \bibinfo{author}{\bibfnamefont{I.~P.} \bibnamefont{Muthuselvam}},
  \bibinfo{author}{\bibfnamefont{T.-R.} \bibnamefont{Chang}},
  \bibinfo{author}{\bibfnamefont{H.~T.} \bibnamefont{Jeng}},
  \bibinfo{author}{\bibfnamefont{G.~S.} \bibnamefont{Murugan}},
  \bibinfo{author}{\bibfnamefont{W.-L.} \bibnamefont{Lee}}, \bibnamefont{and}
  \bibinfo{author}{\bibfnamefont{F.~C.} \bibnamefont{Chou}},
  \bibinfo{journal}{Journal of Physics: Condensed Matter}
  \textbf{\bibinfo{volume}{29}}, \bibinfo{pages}{095601}
  (\bibinfo{year}{2017}).

\bibitem[{\citenamefont{Scharnberg and Klemm}(1980)}]{schar80}
\bibinfo{author}{\bibfnamefont{K.}~\bibnamefont{Scharnberg}} \bibnamefont{and}
  \bibinfo{author}{\bibfnamefont{R.~A.} \bibnamefont{Klemm}},
  \bibinfo{journal}{Phys. Rev. B} \textbf{\bibinfo{volume}{22}},
  \bibinfo{pages}{5233} (\bibinfo{year}{1980}).

\bibitem[{\citenamefont{Clogston}(1962)}]{clogs62}
\bibinfo{author}{\bibfnamefont{A.}~\bibnamefont{Clogston}},
  \bibinfo{journal}{Phys. Rev. Lett.} \textbf{\bibinfo{volume}{9}},
  \bibinfo{pages}{266} (\bibinfo{year}{1962}).

\bibitem[{\citenamefont{Bauer et~al.}(2004)\citenamefont{Bauer, Hilscher,
  Michor, Paul, Scheidt, Gribanov, Seropegin, No\"el, Sigrist, and
  Rogl}}]{bauer04}
\bibinfo{author}{\bibfnamefont{E.}~\bibnamefont{Bauer}},
  \bibinfo{author}{\bibfnamefont{G.}~\bibnamefont{Hilscher}},
  \bibinfo{author}{\bibfnamefont{H.}~\bibnamefont{Michor}},
  \bibinfo{author}{\bibfnamefont{C.}~\bibnamefont{Paul}},
  \bibinfo{author}{\bibfnamefont{E.~W.} \bibnamefont{Scheidt}},
  \bibinfo{author}{\bibfnamefont{A.}~\bibnamefont{Gribanov}},
  \bibinfo{author}{\bibfnamefont{Y.}~\bibnamefont{Seropegin}},
  \bibinfo{author}{\bibfnamefont{H.}~\bibnamefont{No\"el}},
  \bibinfo{author}{\bibfnamefont{M.}~\bibnamefont{Sigrist}}, \bibnamefont{and}
  \bibinfo{author}{\bibfnamefont{P.}~\bibnamefont{Rogl}},
  \bibinfo{journal}{Phys. Rev. Lett.} \textbf{\bibinfo{volume}{92}},
  \bibinfo{pages}{027003} (\bibinfo{year}{2004}).

\bibitem[{\citenamefont{Bay et~al.}(2012)\citenamefont{Bay, Naka, Huang,
  Luigjes, Golden, and de~Visser}}]{bay12}
\bibinfo{author}{\bibfnamefont{T.~V.} \bibnamefont{Bay}},
  \bibinfo{author}{\bibfnamefont{T.}~\bibnamefont{Naka}},
  \bibinfo{author}{\bibfnamefont{Y.~K.} \bibnamefont{Huang}},
  \bibinfo{author}{\bibfnamefont{H.}~\bibnamefont{Luigjes}},
  \bibinfo{author}{\bibfnamefont{M.~S.} \bibnamefont{Golden}},
  \bibnamefont{and}
  \bibinfo{author}{\bibfnamefont{A.}~\bibnamefont{de~Visser}},
  \bibinfo{journal}{Phys. Rev. Lett.} \textbf{\bibinfo{volume}{108}},
  \bibinfo{pages}{057001} (\bibinfo{year}{2012}).

\bibitem[{\citenamefont{Mackenzie et~al.}(1998)\citenamefont{Mackenzie,
  Haselwimmer, Tyler, Lonzarich, Mori, Nishizaki, and Maeno}}]{macke98}
\bibinfo{author}{\bibfnamefont{A.~P.} \bibnamefont{Mackenzie}},
  \bibinfo{author}{\bibfnamefont{R.~K.~W.} \bibnamefont{Haselwimmer}},
  \bibinfo{author}{\bibfnamefont{A.~W.} \bibnamefont{Tyler}},
  \bibinfo{author}{\bibfnamefont{G.~G.} \bibnamefont{Lonzarich}},
  \bibinfo{author}{\bibfnamefont{Y.}~\bibnamefont{Mori}},
  \bibinfo{author}{\bibfnamefont{S.}~\bibnamefont{Nishizaki}},
  \bibnamefont{and} \bibinfo{author}{\bibfnamefont{Y.}~\bibnamefont{Maeno}},
  \bibinfo{journal}{Phys. Rev. Lett.} \textbf{\bibinfo{volume}{80}},
  \bibinfo{pages}{161} (\bibinfo{year}{1998}).

\bibitem[{\citenamefont{Abrikosov and Gor'kov}(1961)}]{abrik61}
\bibinfo{author}{\bibfnamefont{A.~A.} \bibnamefont{Abrikosov}}
  \bibnamefont{and} \bibinfo{author}{\bibfnamefont{L.~P.}
  \bibnamefont{Gor'kov}}, \bibinfo{journal}{Sov. Phys. JETP}
  \textbf{\bibinfo{volume}{12}}, \bibinfo{pages}{1243} (\bibinfo{year}{1961}).

\bibitem[{\citenamefont{Xu et~al.}(2019)\citenamefont{Xu, Kang, Chang, Lin,
  Bian, Yuan, Qu, Zhang, and Jia}}]{xu19}
\bibinfo{author}{\bibfnamefont{X.}~\bibnamefont{Xu}},
  \bibinfo{author}{\bibfnamefont{Z.}~\bibnamefont{Kang}},
  \bibinfo{author}{\bibfnamefont{T.-R.} \bibnamefont{Chang}},
  \bibinfo{author}{\bibfnamefont{H.}~\bibnamefont{Lin}},
  \bibinfo{author}{\bibfnamefont{G.}~\bibnamefont{Bian}},
  \bibinfo{author}{\bibfnamefont{Z.}~\bibnamefont{Yuan}},
  \bibinfo{author}{\bibfnamefont{Z.}~\bibnamefont{Qu}},
  \bibinfo{author}{\bibfnamefont{J.}~\bibnamefont{Zhang}}, \bibnamefont{and}
  \bibinfo{author}{\bibfnamefont{S.}~\bibnamefont{Jia}},
  \bibinfo{journal}{Phys. Rev. B} \textbf{\bibinfo{volume}{99}},
  \bibinfo{pages}{104516} (\bibinfo{year}{2019}).

\bibitem[{\citenamefont{Sasaki et~al.}(2011)\citenamefont{Sasaki, Kriener,
  Segawa, Yada, Tanaka, Sato, and Ando}}]{sasak11}
\bibinfo{author}{\bibfnamefont{S.}~\bibnamefont{Sasaki}},
  \bibinfo{author}{\bibfnamefont{M.}~\bibnamefont{Kriener}},
  \bibinfo{author}{\bibfnamefont{K.}~\bibnamefont{Segawa}},
  \bibinfo{author}{\bibfnamefont{K.}~\bibnamefont{Yada}},
  \bibinfo{author}{\bibfnamefont{Y.}~\bibnamefont{Tanaka}},
  \bibinfo{author}{\bibfnamefont{M.}~\bibnamefont{Sato}}, \bibnamefont{and}
  \bibinfo{author}{\bibfnamefont{Y.}~\bibnamefont{Ando}},
  \bibinfo{journal}{Phys. Rev. Lett.} \textbf{\bibinfo{volume}{107}},
  \bibinfo{pages}{217001} (\bibinfo{year}{2011}).

\bibitem[{\citenamefont{Michaeli and Fu}(2012)}]{micha12}
\bibinfo{author}{\bibfnamefont{K.}~\bibnamefont{Michaeli}} \bibnamefont{and}
  \bibinfo{author}{\bibfnamefont{L.}~\bibnamefont{Fu}}, \bibinfo{journal}{Phys.
  Rev. Lett.} \textbf{\bibinfo{volume}{109}}, \bibinfo{pages}{187003}
  (\bibinfo{year}{2012}).

\bibitem[{\citenamefont{Nagai et~al.}(2014)\citenamefont{Nagai, Ota, and
  Machida}}]{nagai14}
\bibinfo{author}{\bibfnamefont{Y.}~\bibnamefont{Nagai}},
  \bibinfo{author}{\bibfnamefont{Y.}~\bibnamefont{Ota}}, \bibnamefont{and}
  \bibinfo{author}{\bibfnamefont{M.}~\bibnamefont{Machida}},
  \bibinfo{journal}{Phys. Rev. B} \textbf{\bibinfo{volume}{89}},
  \bibinfo{pages}{214506} (\bibinfo{year}{2014}).

\bibitem[{\citenamefont{Matano et~al.}(2016)\citenamefont{Matano, Kriener,
  Segawa, Ando, and Zheng}}]{matan16}
\bibinfo{author}{\bibfnamefont{K.}~\bibnamefont{Matano}},
  \bibinfo{author}{\bibfnamefont{M.}~\bibnamefont{Kriener}},
  \bibinfo{author}{\bibfnamefont{K.}~\bibnamefont{Segawa}},
  \bibinfo{author}{\bibfnamefont{Y.}~\bibnamefont{Ando}}, \bibnamefont{and}
  \bibinfo{author}{\bibfnamefont{G.-q.} \bibnamefont{Zheng}},
  \bibinfo{journal}{Nat Phys} \textbf{\bibinfo{volume}{12}},
  \bibinfo{pages}{852} (\bibinfo{year}{2016}).

\bibitem[{\citenamefont{Pan et~al.}(2016)\citenamefont{Pan, Nikitin, Araizi,
  Huang, Matsushita, Naka, and de~Visser}}]{pan16}
\bibinfo{author}{\bibfnamefont{Y.}~\bibnamefont{Pan}},
  \bibinfo{author}{\bibfnamefont{A.~M.} \bibnamefont{Nikitin}},
  \bibinfo{author}{\bibfnamefont{G.~K.} \bibnamefont{Araizi}},
  \bibinfo{author}{\bibfnamefont{Y.~K.} \bibnamefont{Huang}},
  \bibinfo{author}{\bibfnamefont{Y.}~\bibnamefont{Matsushita}},
  \bibinfo{author}{\bibfnamefont{T.}~\bibnamefont{Naka}}, \bibnamefont{and}
  \bibinfo{author}{\bibfnamefont{A.}~\bibnamefont{de~Visser}},
  \bibinfo{journal}{Scientific Reports} \textbf{\bibinfo{volume}{6}},
  \bibinfo{pages}{28632 EP } (\bibinfo{year}{2016}).

\bibitem[{\citenamefont{Smylie et~al.}(2017)\citenamefont{Smylie, Willa, Claus,
  Snezhko, Martin, Kwok, Qiu, Hor, Bokari, Niraula et~al.}}]{smyli17}
\bibinfo{author}{\bibfnamefont{M.~P.} \bibnamefont{Smylie}},
  \bibinfo{author}{\bibfnamefont{K.}~\bibnamefont{Willa}},
  \bibinfo{author}{\bibfnamefont{H.}~\bibnamefont{Claus}},
  \bibinfo{author}{\bibfnamefont{A.}~\bibnamefont{Snezhko}},
  \bibinfo{author}{\bibfnamefont{I.}~\bibnamefont{Martin}},
  \bibinfo{author}{\bibfnamefont{W.-K.} \bibnamefont{Kwok}},
  \bibinfo{author}{\bibfnamefont{Y.}~\bibnamefont{Qiu}},
  \bibinfo{author}{\bibfnamefont{Y.~S.} \bibnamefont{Hor}},
  \bibinfo{author}{\bibfnamefont{E.}~\bibnamefont{Bokari}},
  \bibinfo{author}{\bibfnamefont{P.}~\bibnamefont{Niraula}},
  \bibnamefont{et~al.}, \bibinfo{journal}{Phys. Rev. B}
  \textbf{\bibinfo{volume}{96}}, \bibinfo{pages}{115145}
  (\bibinfo{year}{2017}).

\bibitem[{\citenamefont{Yonezawa et~al.}(2017)\citenamefont{Yonezawa, Tajiri,
  Nakata, Nagai, Wang, Segawa, Ando, and Maeno}}]{yonez17}
\bibinfo{author}{\bibfnamefont{S.}~\bibnamefont{Yonezawa}},
  \bibinfo{author}{\bibfnamefont{K.}~\bibnamefont{Tajiri}},
  \bibinfo{author}{\bibfnamefont{S.}~\bibnamefont{Nakata}},
  \bibinfo{author}{\bibfnamefont{Y.}~\bibnamefont{Nagai}},
  \bibinfo{author}{\bibfnamefont{Z.}~\bibnamefont{Wang}},
  \bibinfo{author}{\bibfnamefont{K.}~\bibnamefont{Segawa}},
  \bibinfo{author}{\bibfnamefont{Y.}~\bibnamefont{Ando}}, \bibnamefont{and}
  \bibinfo{author}{\bibfnamefont{Y.}~\bibnamefont{Maeno}},
  \bibinfo{journal}{Nat Phys} \textbf{\bibinfo{volume}{13}},
  \bibinfo{pages}{123} (\bibinfo{year}{2017}).

\bibitem[{\citenamefont{Le et~al.}(2020)\citenamefont{Le, Sun, Jin, Che, Yin,
  Li, Pang, Xu, Zhao, Kittaka et~al.}}]{le20}
\bibinfo{author}{\bibfnamefont{T.}~\bibnamefont{Le}},
  \bibinfo{author}{\bibfnamefont{Y.}~\bibnamefont{Sun}},
  \bibinfo{author}{\bibfnamefont{H.-K.} \bibnamefont{Jin}},
  \bibinfo{author}{\bibfnamefont{L.}~\bibnamefont{Che}},
  \bibinfo{author}{\bibfnamefont{L.}~\bibnamefont{Yin}},
  \bibinfo{author}{\bibfnamefont{J.}~\bibnamefont{Li}},
  \bibinfo{author}{\bibfnamefont{G.}~\bibnamefont{Pang}},
  \bibinfo{author}{\bibfnamefont{C.}~\bibnamefont{Xu}},
  \bibinfo{author}{\bibfnamefont{L.}~\bibnamefont{Zhao}},
  \bibinfo{author}{\bibfnamefont{S.}~\bibnamefont{Kittaka}},
  \bibnamefont{et~al.}, \bibinfo{journal}{Science Bulletin}
  \textbf{\bibinfo{volume}{65}}, \bibinfo{pages}{1349 } (\bibinfo{year}{2020}).

\bibitem[{\citenamefont{Shi et~al.}(2003)\citenamefont{Shi, Tokunaga, Tamegai,
  Takano, Togano, Kito, and Ihara}}]{shi03}
\bibinfo{author}{\bibfnamefont{Z.~X.} \bibnamefont{Shi}},
  \bibinfo{author}{\bibfnamefont{M.}~\bibnamefont{Tokunaga}},
  \bibinfo{author}{\bibfnamefont{T.}~\bibnamefont{Tamegai}},
  \bibinfo{author}{\bibfnamefont{Y.}~\bibnamefont{Takano}},
  \bibinfo{author}{\bibfnamefont{K.}~\bibnamefont{Togano}},
  \bibinfo{author}{\bibfnamefont{H.}~\bibnamefont{Kito}}, \bibnamefont{and}
  \bibinfo{author}{\bibfnamefont{H.}~\bibnamefont{Ihara}},
  \bibinfo{journal}{Phys. Rev. B} \textbf{\bibinfo{volume}{68}},
  \bibinfo{pages}{104513} (\bibinfo{year}{2003}).

\bibitem[{\citenamefont{Zhu et~al.}(2010)\citenamefont{Zhu, Lu, Sun, Pi, Qu,
  Ling, Yang, and Zhang}}]{zhu10a}
\bibinfo{author}{\bibfnamefont{X.~D.} \bibnamefont{Zhu}},
  \bibinfo{author}{\bibfnamefont{J.~C.} \bibnamefont{Lu}},
  \bibinfo{author}{\bibfnamefont{Y.~P.} \bibnamefont{Sun}},
  \bibinfo{author}{\bibfnamefont{L.}~\bibnamefont{Pi}},
  \bibinfo{author}{\bibfnamefont{Z.}~\bibnamefont{Qu}},
  \bibinfo{author}{\bibfnamefont{L.~S.} \bibnamefont{Ling}},
  \bibinfo{author}{\bibfnamefont{Z.~R.} \bibnamefont{Yang}}, \bibnamefont{and}
  \bibinfo{author}{\bibfnamefont{Y.~H.} \bibnamefont{Zhang}},
  \bibinfo{journal}{Journal of Physics: Condensed Matter}
  \textbf{\bibinfo{volume}{22}}, \bibinfo{pages}{505704}
  (\bibinfo{year}{2010}).

\end{thebibliography}
\end{document}